\documentstyle[pra,aps,graphicx]{revtex}
\begin{document}

\title{Second magnetization peak in flux lattices - the decoupling scenario}
\author{Baruch Horovitz} 
\address {Department of Physics, Ben-Gurion 
University of the Negev, Beer-Sheva 84105, Israel} 
\maketitle 
\widetext
\begin{abstract}
The second peak phenomena of flux lattices in layered superconductors 
is described in terms of a disorder induced layer decoupling transition. 
For weak disorder the tilt mudulus undergoes an apparent discontinuity 
which leads to an enhanced critical current
and reduced domain size in the decoupled phase. 
The Josephson plasma frequency is reduced by decoupling and by  
Josephson glass pinning; in the liquid phase it varies as 
$1/[BT(T+T_0)]$ where $T$ is temperature, $B$ is field and $T_0$ is the 
disorder dependent temperature of the multicritical point.
\end{abstract}


Vortex matter in the presence of disorder has emerged as a fundamental 
problem of elastic manifolds in a random media \cite{Blatter}. 
Impurity disorder does not allow long range translational order of 
the flux lattice and finite domains are expected \cite{LO}.
At low temperatures and fields the system is a Bragg glass 
\cite{Giamarchi,Natterman}, i.e. the lattice is dislocation free, 
at long scales the displacement 
correlations decay as a power law and Bragg peaks are expected.
The impurity induced domains are essential for the description of 
both equilibrium, e.g. thermodynamic phase transitions
and non-equilibrium, e.g. critical current phenomena. 

The critical current $j_c$ measures the pinning force in  
the domains \cite{Blatter,LO}. Increasing the magnetic field or temperature 
reduces the pinning force and $j_c$ is decreased. However, in many 
type II superconductors a sharp enhancement of  $j_c$ is observed at a 
``second peak'' field $B_0$. This peak phenomena is 
most pronounced in layered superconductors such as 
$Bi_2Sr_2CaCu_2O_8$ (BSSCO) \cite{Kes,Khaykovich1,Khaykovich2}, 
$YBa_{2}Cu_{3}O_{7}$ (YBCO) \cite{Deligiannis}, in $NbSe_2$ 
\cite{Higgins,Marley} and in $Pb/Ge$ multilayers \cite{Bruynseraede} 
for fields perpendicular to the layers. The second 
peak phenomena signals that pinning becomes more effective, e.g. 
due to softening of the flux lattice \cite{LO}. The reason for 
softening could be the approach to melting \cite{Otterlo}; however, 
neutron scattering data on BSCCO \cite{Forgan} shows that Bragg spots 
of the flux lattice persist well above $B_0$.

Disorder plays an essential role also in the equilibrium phase diagram 
of layered 
superconductors. This has been most extensively studied in BSSCO 
\cite{Kes,Khaykovich1,Khaykovich2,Fuchs1,Fuchs2}. The second peak  
corresponds to a phase transition \cite{Khaykovich2} 
in the range $500-900G$ (decreasing with disorder) and is weakly 
temperature dependent up to a temperature $T_0 \approx 40 K$. The 
point $B_0$, $T_0$ is a multicritical point where the second peak 
transition meets a first order transition as well as two depinning 
lines. Thus the second peak manifests both equilibrium and 
nonequilibrium phenomena of disorder in flux lattices and its 
understanding presents a fundamental challenge.

For a flux lattice with point impurities, by using renormalization 
group (RG) and replica symmetry breaking (RSB) methods we have 
derived \cite{Horovitz} a phase diagram with four phases, which all 
meet at a multicritical point $B_0$, $T_0$, in remarkable 
correspondence with data on BSCCO. 
The present work focuses on the layer decoupling transition at a 
temperature independent field $B_0$ for $T<T_0$. 
As shown here, the fusion of Bragg 
glass concepts with decoupling accounts for the peculiar second peak 
phenomena, i.e. the enhanced $j_c$. The Josephson plasma resonance 
is also considered as a probe of the Josephson coupling 
\cite{Matsuda,Koshelev}, being reduced 
by decoupling and by a Josephson glass parameter. 
Very recent data on BSCCO has indeed shown a significant reduction in the
resonance frequency at the second peak transition
\cite{Shibauchi,Matsuda2}.

It has been recently shown that decoupling coalesces with a
defect unbinding transition \cite{Dodgson} which has analogs in isotropic
systems \cite{Frey}. The resulting vacancies and interstitials lead to a
reduction in the elastic tilt mudulus \cite{Marchetti}, consistent with
the decoupling scenario as described below.
 It is possible then that a decoupling-defect
transition accounts for the peak phenomena in all type II superconductors.
The analysis below is, however, presented for layered anisotropic systems
where quantitative predictions can be made.

In a layered superconductor each flux line is composed of one point 
singularity, or a pancake vortex, in each layer. When the pancake 
vortices fluctuate they can generate a divergence in the Josephson 
phase, leading to a renormalized Josephson coupling $E_J^R$ which 
vanishes in the decoupled $B>B_0$ phase \cite{Horovitz,Glazman,Daemen}. 
The 3-dimensional flux lattice is 
still present in the decoupled phase (in the Bragg glass sense), 
with interlayer coupling mediated by the magnetic field. Before 
presenting a microscopic model, I start with a rather simple description 
of elasticity within domains, which shows the 
second peak transition, i.e. $j_c$ enhancement at decoupling.

The transverse tilt modulus of a flux lattice in a layered 
superconductor for fields perpendicular to the layers is given by 
\cite{Sudbo,GK,Goldin}
\begin{equation}
c_{44}({\bf q},k) = \frac{\tau}{32\pi \lambda_{ab}^2 d} + \frac{B^2 }{4\pi} 
\frac{1}{1+\lambda_c^2 q^2 + \lambda_{ab}^2 k^2} +\frac{2B\phi_0}{(8\pi
\lambda_c)^2} \ln (a^2/4\pi \xi_0 ^2)  \label{c44}
\end{equation}
where ${\bf q}$ and $k$ are momenta parallel and perpendicular to the 
layers, respectively, $\lambda_{ab}$ and $\lambda_c$ are the London 
penetration lengths parallel and perpendicular to the layers, 
respectively, $\phi_0$ is the flux quantum, $a^2=\phi_0/B$ is the 
unit cell area, d is the interlayer spacing, $\xi_0$ is the 
in-layer coherence length and $\tau = \phi_0^2 
d/(4\pi^2\lambda_{ab}^2)$ sets the energy scale. The first term of 
Eq.\ (\ref{c44}) is due to the magnetic coupling, while the 2nd and 3rd terms 
originate from the Josephson coupling energy per unit area $E_J$, i.e.
$\lambda_c^2=\tau\lambda_{ab}^2/(4\pi E_Jd^2)$. 
 The second term is peculiar: at 
$q\neq 0$ it vanishes when $E_J$ vanishes and 
$\lambda_c \rightarrow \infty$, as it should. However, at $q=0$ this 
term seems to survive even if $\lambda_c \rightarrow \infty$. The 
origin of this peculiarity is that the harmonic expansion of the 
Josephson cosine term which identifies $c_{44}$ fails\cite{Goldin}
when both $q, \,1/\lambda_c \rightarrow 0$. In fact, the nonlinear 
cosine term generates a renormalized $\lambda_c^R$ which diverges at 
decoupling.

The Bragg glass domain size $R_{BG}$ (parallel to the layers) sets a 
scale for the relevant $q$ values. When $R_{BG} > \lambda_c^R$ the 
tilt mudulus is large, containing the $B^2/4\pi$ term of Eq.\ 
(\ref{c44}). However, as decoupling at the field $B_0$ is approached 
$\lambda_c^R$ diverges and when $R_{BG} < \lambda_c^R \,$ Eq.\ (\ref{c44}) fails 
to describe $c_{44}$ on the scale of $q\approx 1/R_{BG}$. This defines 
an anharmonic crossover regime where usual elasticity cannot be used to derive 
Bragg glass properties. Finally, at $B>B_0$ elasticity is restored 
and $c_{44}$ is reduced to the first term in Eq.\ (\ref{c44}). The main 
interest is in the regime of strong fields, i.e. $a\lesssim 
2\lambda_{ab}$ where $T_0<\tau$ is below 
melting \cite{Horovitz}. Thus at $B<B_0$ and for 
sufficiently large domains the second term in Eq. (1) dominates and 
$c_{44}$ has an apparent discontinuity,
\begin {mathletters}
\begin{eqnarray}
c_{44} &=& \pi \lambda_{ab}^2\tau/da^4    \hspace{18mm} \lambda_c^R < 
R_{BG} \label{c44a}\\
c_{44} &=& \tau/(32\pi \lambda_{ab}^2 d)   \hspace{15mm}  \lambda_c^R = 
\infty  \label{c44b}
\end{eqnarray}
\end{mathletters}
Hence $c_{44}$ is reduced within the anharmonic regime by the small 
factor $\epsilon=a^4/(32\pi^2\lambda_{ab}^4)$. 

The apparent discontinuity in $c_{44}$ affects also the domain sizes 
which can be estimated by a dimensional argument \cite{LO,Giamarchi}. Consider 
the tilt $c_{44}$ and shear $c_{66}$ terms of the elasticity 
Hamiltonian for the displacement ${\bf u}({\bf r})$ and its
transverse component ${\bf u}_T({\bf r})$. 
Rescaling parallel and perpendicular lengths yields an isotropic form
 \cite{Blatter,Natterman}
\begin{equation}
{\cal H}= 
\int d^3r\{\case{1}{2} c_{44}^{1/3}c_{66}^{2/3}
[{\bf \nabla}u_T({\bf r})]^2
-(\xi_0^2/a^2d)U_{pin}({\bf r})
\sum_{{\bf Q}}\cos {\bf Q}\cdot 
[{\mbox{\boldmath $\rho$}}- {\bf u} ({\bf r})]\} \label{H}
\end{equation}
where $U_{pin}({\bf r})$ is a random potential in 3-dimensional ${\bf 
r}=({\mbox{\boldmath $\rho$}}, z)$ which couples to the flux density
modulations with wavevectors 
${\bf Q}$; its disorder average is
$\langle U_{pin}({\bf r})U_{pin}({\bf r}')\rangle =\case{1}{2}
d{\bar U}\delta ^3 ({\bf r}-{\bf r}')$.
Disorder average over configurations
 ${\bf u} ({\bf r})$ and ${\bf u}' ({\bf r})$
yields $\sum_{{\bf Q}}\cos {\bf Q}\cdot[{\bf u} ({\bf r})-
{\bf u}' ({\bf r})]$; the sum is cutoff by $Q \lesssim \langle
u_T^2 \rangle ^{-1/2}$ where $\langle u^2 \rangle \approx \langle u_T^2
\rangle$ are the fluctuations
in a domain of size $R'$. Thus averaging  Eq.\ (\ref{H})  yields
\begin{equation}
\langle H \rangle /R'^3 = \case{1}{2}c_{44}^{1/3}c_{66}^{2/3} 
\langle u_T^2 \rangle R'^{-2} - {\bar U}^{1/2}\xi_0^2
/[a^2d\langle u_T^2 \rangle R'^3]^{1/2} \, .    \label{Hav}
\end{equation}
Minimizing with respect to $R'$ yields $R' \sim \langle u_T^2 \rangle ^3$, 
i.e. the Flory exponent \cite{Giamarchi}. The domain size 
parallel to the layers is (up to $\ln(a/d)$ and a numerical prefactor)
\begin{eqnarray}
R & \approx & (\lambda_{ab}/a)^5 \langle u_T^2 \rangle ^3 /(s\xi_0^4 d)
 \hspace{20mm} \lambda_c^R < R 
\nonumber\\
R  & \approx & (\lambda_{ab}/a)^3\langle u_T^2 \rangle ^3
/(4\pi s\xi_0^4d)
\hspace{15mm}  \lambda_c^R = \infty  \label{R}
\end{eqnarray}
where $c_{66}=\tau/16da^2$ \cite{Sudbo,GK,Goldin},
 $s=4\pi {\bar p}{\bar U}\lambda_{ab}^4/[\tau^2a^2\ln ^2
(a/d)]$ defines the decoupling transition at $s=\case{1}{2}$ and ${\bar
p}\approx 1$ is defined below.

The pinning length $R=R_p$ is given by Eq.\ (\ref{R}) with 
$\langle u_T^2 \rangle \approx \xi_0^2$. To allow for large pinning
domains one needs either $a<\lambda_{ab}$ or to allow for domains with
a somewhat larger fluctuations in $\langle u_T^2 \rangle$; the latter
increases $R_p$ very rapidly  since it
increases with the 6-th power of $u_T$. 
The critical current can now 
be estimated \cite{Blatter,LO} by balancing the Lorenz force 
$j_cBR^3/c$ with the pinning force $\langle H \rangle /\xi_0$ 
(evaluated at the minimum of Eq.\ (\ref{Hav})), leading to $j_c\sim 
1/c_{44}$. Increasing the field within the 
anharmonic regime decreases $c_{44}$ by the factor $\epsilon$ so 
that $j_c$ is significantly enhanced when $a\lesssim \lambda_{ab}$.
Note that the measured magnetization changes (and inferred $j_c$) at 
$B_0$ decrease with temperature due to the strongly temperature 
dependent relaxation rates \cite{Yeshurun}, approaching the much 
smaller equilibrium magnetizations.

A second length scale $R=R_{BG}$ is identified by  Eq.\ (\ref{R})
with the fluctuations $\langle u_T^2 \rangle \approx a^2$.
 The proper definition of $R_{BG}$ is the scale for the onset of the
 $\ln r$ form 
for the displacement correlation function. While the derivation from 
Eq.\ (\ref{Hav}) cannot capture this $\ln r$, it does give the right 
expression for $R_{BG}$ \cite {Giamarchi}. Thus, $R_{BG}$ depends on 
$c_{44}$ and is reduced by $\epsilon ^{1/2}$ through the anharmonic regime. 
The latter depends also on $\lambda_c^R$ for which $\ln \lambda_c^R \sim
(B-B_0)^{-1}$ in the RSB or 1st order RG solutions \cite{Horovitz}, though 
$\ln \lambda_c^R \sim(B-B_0)^{-1/2}$ in second order RG 
\cite{Horovitz2}; decoupling 
may also be of first order \cite{Daemen}, leading to a narrower 
anharmonic regime. Fig. 1 illustrates the 
lengths $R_{BG}$ and $\lambda_c^R$, demonstrating the anharmonic 
regime within which $R_{BG}$ has a significant drop and 
correspondingly $j_c$ has an apparent jump.
Note that even in the decoupled phase ($B>B_0$) $R_{BG}$ is large
for typical type II superconductors, 
 $R_{BG}\approx \lambda_{ab}^3 a^3 /(4\pi s \xi_0^4 d) \gg a$, consistent
with a decoupling transition within the Bragg glass phase, i.e. below a
melting transition.

\newpage
\begin{figure}[htb]
\begin{center}
\includegraphics[scale=0.7]{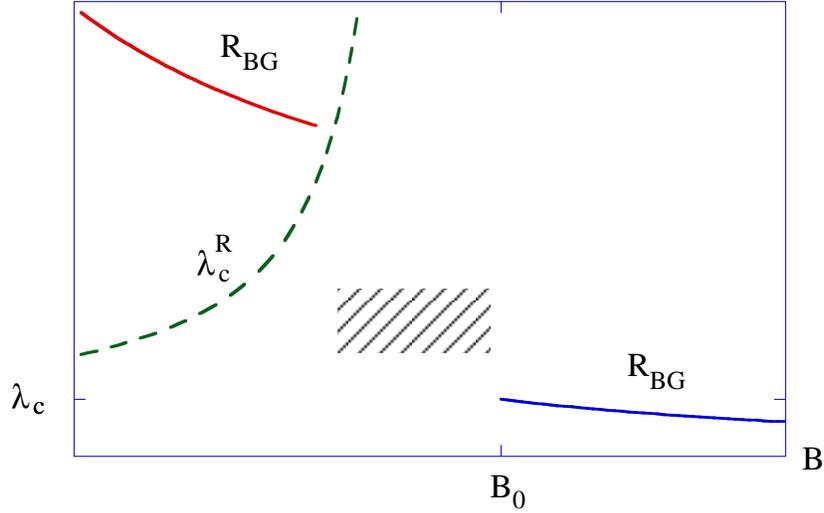}
\end{center}
\caption{ Bragg glass domain size $R_{BG}$ parallel to the layers
and the renormalized London length perpendicular to the layers  
$\lambda_c^R$; the latter diverges at the decoupling field $B_0$.
$R_{BG}$ can be found from elasticity for $B<B_0$ only if   
$R_{BG}>\lambda_c^R$; otherwise, as in the hatched region,
the elastic tilt mudulus is ill defined. }
\end{figure}
\hspace{15mm}

I proceed now to derive the lattice displacement correlation allowing
for a renormalized Josephson coupling and for a Josephson glass order 
parameter. This derivation avoids the harmonic expansion for the 
elastic modulii and shows how the Bragg glass domain sizes are directly 
affected by the renormalized $\lambda_c^R$.
The Josephson phase between the layers $n$ and $n+1$ at position ${\bf 
r}$ in the layer involves contributions from a nonsingular component 
$\theta_n({\bf r})$ and from singular vortex terms \cite{Horovitz2}. 
Consider a flux lattice with an equilibrium position of the $l$-th 
flux line at ${\bf R}_{l}$. The singular 
phase around a pancake vortex at position ${\bf R}_{l} + {\bf u}_{l}^{n}$ 
is $\alpha ({\bf r} - {\bf R}_{l} - {\bf u}_{l}^{n})$ where $\alpha 
({\bf r}) = \arctan (y/x)$ with ${\bf r} = (x,y)$. Expansion of the 
interlayer phase difference $\alpha ({\bf r} - {\bf R}_{l} - 
{\bf u}_{l}^{n}) - \alpha ({\bf r} - {\bf R}_{l} - {\bf u}_{l}^{n+1})$
yields for the singular part of the Josephson phase $b_{n}({\bf r}) = 
\sum_{l}({\bf u}_{l}^{n+1} - {\bf u}_{l}^{n}){\bf 
\nabla} \alpha ({\bf r} - {\bf R}_{l})$. The Hamiltonian for the 
transverse displacements involves also the magnetic contributions to the 
shear modulus $c_{66}= \tau/(16da^{2})$ and the tilt mudulus
\[
c_{44}^{0}(k) = [\tau/(8da^{2}\lambda _{ab}^{2}k_{z}^{2})] \ln 
(1+a^{2}k_{z}^{2}/4\pi)
\]
 where $k_z=(2/d)\sin (kd/2)$; its $k\rightarrow 0$ form is the first term
in Eq.\ (\ref{c44}).
This leads to the Hamiltonian of the pure system
\begin {equation}
{\cal H}_{pure}/T = \case{1}{2}\sum_{q,k} G_{f}^{-1}({\bf q},k) 
|\theta({\bf q},k)|^{2}  +
\case{1}{2} \sum_{{\bf q},k} c({\bf q},k)q^2|b({\bf q},k)|^{2}
-\frac{E_J}{T} \sum_n \int d^{2}r \, \cos [\theta_n({\bf r}) + b_{n}({\bf r})]
\label{Hpure}\end{equation}

Here $c({\bf q},k)=(a^2/2\pi d)^2
[k_z^2c_{44}^{0}(k)+q^2c_{66}]/Tk_z^2$, $E_J$ is the Josephson coupling
and the coefficient of the non-singular phase is \cite{Horovitz2}
$G_{f}(q,k)=4\pi d^3T(\lambda_{ab}^{-2}+k_z^2)/(\tau q^2)$.
The conventional $c_{44}$ is obtained by expanding the cosine term in 
Eq.\ (\ref{Hpure}) and shifting $\theta({\bf q},k)$ to eliminate the cross term. 
The latter shift leads to an expansion parameter \cite{Goldin} with 
terms $\sim q^2k_z^2|u_{T}({\bf q},k)|^2/
[q^2+\lambda_c^{-2}(1+\lambda_{ab}^2k_z^2)]^2$, i.e. these diverge when both 
$q,1/\lambda_c \rightarrow 0$ and the expansion becomes invalid. 

Consider now a pinning potential $U_{pin}^{n}({\bf r})$ which couples 
to the vortex shape function $p({\bf r})$ leading to a pinning energy
$\int d^{2}r \sum_{n,l} U_{pin}^{n}({\bf r})
p({\bf r} - {\bf R}_{l} - {\bf u}_{l}^{n})$ .
The aim is to identify domain sizes $R_p$ (and infer $R_{BG}$), hence 
the pinning energy is expanded in
 ${\bf u}_{l}^{n}$ and a replica average with 
the weight $\exp\{-\int d^2r\sum_n[U_{pin}^n({\bf r})]^2/\bar{U}\}$ 
then leads to $\exp[(\bar{U}\bar{p}/4T^2)\sum_{n,l}\sum_{\alpha,\beta}
{\bf u}_{l}^{n,\alpha}\cdot {\bf u}_{l}^{n,\beta}]$ where $\int 
\partial_ip({\bf r})\partial_jp({\bf r})d^2r=\bar{p}\delta_{i,j}$
and $\alpha,\beta=1,2,\ldots,n$ are replica indices. 

The $b^{\alpha}({\bf q},k)$ variables can be decoupled from the total 
Josephson phase $\tilde{b}_n({\bf r})=b_n({\bf r})+\theta_n({\bf r})$
by shifting to
$d^{\alpha}({\bf q},k)=b^{\alpha}({\bf q},k)-B_{\gamma,\alpha}({\bf q},k)
G_{f}^{-1}({\bf q},k)\tilde{b}^{\alpha}({\bf q},k)$ where 
\[B^{-1}_{\alpha,\beta}({\bf q},k)=G_{f}^{-1}({\bf q},k)\alpha({\bf q},k)
\delta_{\alpha,\beta} - s_0q^2/k_z^2 \, ,
\]
 $\alpha({\bf q},k)=1+G_f({\bf q},k)c({\bf q},k)q^2$
and  $s_0=\bar{U}\bar{p}a^2d/(4\pi d^2T)^2$.
The resulting replicated Hamiltonian is 
\begin{eqnarray}
{\cal H}_r &=& \case{1}{2}\sum_{q,k;\alpha,\beta} B_{\alpha,\beta}^{-1}
d^{\alpha}({\bf q},k)d^{\beta *}({\bf q},k)
+\case{1}{2}[c({\bf q},k)\alpha^{-1}({\bf q},k)q^2\delta_{\alpha,\beta} - 
s_0\alpha^{-2}({\bf q},k)q^2/k_z^2]\tilde{b}^{\alpha}({\bf q},k)
\tilde{b}^{\beta *}({\bf q},k) \nonumber\\
&-&\frac{E_J}{T}\sum_{n;\alpha}\int d^2r\cos \tilde{b}_n^{\alpha}({\bf r})
-\frac{E_v}{T}\sum_{n;\alpha\neq \beta}\int d^2r\cos 
[\tilde{b}_n^{\alpha}({\bf r})-\tilde{b}_n^{\beta}({\bf r})]
\end{eqnarray}
 The inter-replica 
$E_v$ term is generated from the Josephson coupling in second order 
RG. It is essential to keep it from the start since it generates a 
Josephson glass parameter and affects the value of the decoupling 
field \cite{Horovitz}. 

The $\alpha({\bf q},k)$ factor, which results from the nonsingular 
phase, is for $\epsilon \ll 1$ very close to $1$ for all 
${\bf q},k$ values except when both $k<1/\lambda_{ab}$ and 
$q>ka/\lambda_{ab}$. The phase transitions are dominated
by $k>1/a$ modes so that our previous phase diagram is 
recovered (\onlinecite{Horovitz} Fig.1). In particular there is a 
multicritical point at a field $B_0$ where $s=\case{1}{2}$
and temperature $T_0=\tau a^2\ln (a/d)/8\pi\lambda_{ab}^2$. At 
$B=B_0$ and $T<T_0$ there is a 
decoupling transition at which the renormalized Josephson coupling 
$z$ (with bare value $z_{bare}=E_J/Td$) vanishes. Note that 
the higher $B_0$ of YBCO as compared to BSCCO is consistent with a
shorter $\lambda_{ab}$ and a somewhat weaker disorder.

The fluctuations in $u_T({\bf q},k)$ in terms of the shifted 
variables, using the RSB solution \cite{Horovitz} 
are given by
\begin{eqnarray}
\langle |u_T({\bf q},k)|^2\rangle &=& (2\pi d^2)^{-2}\frac{q^2}{k_z^4}
[s_0q^2 G_f({\bf q},k)\alpha^{-1}({\bf q},k) \left(c({\bf q},k)
q^2 + \frac{G_f^{-1}({\bf q},k)z}{G_f^{-1}({\bf 
q},k)+z}\right)^{-1}\nonumber\\
&+& \frac{s_0}{c({\bf q},k)\alpha^2({\bf q},k)}\left(\frac{c({\bf q},k)}
{\alpha({\bf q},k)}q^2 + z\right)^{-1}] + \ldots  \label{fluc}
\end{eqnarray}
where $\ldots$ stands for terms which converge in $({\bf q},k)$ 
integration. Note the term $G_f^{-1}({\bf q},k)z/[G_f^{-1}({\bf 
q},k)+z]$ which depends on the order of $q\rightarrow 0$ and 
$z\rightarrow 0$ limits; this limit dependence leads to the 
apparent discontinuity in $c_{44}$ as discussed above. For $z\neq 0$ 
and small $q$, i.e. $G_f^{-1}({\bf q},k) \ll z$ the first term in 
Eq.\ (\ref{fluc}) dominates, leading to
\begin{equation}
\langle |u_T({\bf q},k)|^2\rangle \approx \frac{4\pi^2s_0T^2}
{a^8[c_{44}k^2 + c_{66}q^2]^2}  \hspace{10mm} q<1/\lambda_c^R
\end{equation}
where $c_{44}$ is from Eq.\ (\ref{c44a}) and the condition 
$G_f^{-1}({\bf q},k) \ll z$ is written in terms of a renormalized 
London length $\lambda_c^R=[\lambda_{ab}^2\tau/(4\pi Td^3z)]^{1/2}$. 
The correlations at distance $r$ parallel to the layers are then
\begin{equation}
\langle [u_T(r)-u_T(0)]^2\rangle 
\approx \frac{4d^2s_0T^2}{a^4c_{44}^{1/2}c_{66}^{3/2}}r
\equiv \xi_0^2\frac{r}{R_p} \label{Rp1}\, .
\end{equation}
The last equality defines the pinning length $R_p$ where the 
fluctuations become of order $\xi_0^2$. This result for $R_p$ (up to a 
numerical prefactor) is 
 the same as the one obtained from Eq.\ (\ref{R}) with
 $\langle u_T^2 \rangle \approx a^2$. The Bragg glass domain 
size is enhanced by $R_{BG}\approx R_p(a/\xi_0)^6$, as discussed above.

In the decoupled phase with $z=0$ the second term in Eq.\ 
(\ref{fluc}) dominates. To leading order in $\epsilon$ the result is 
identical to Eq.\ (\ref{Rp1}) except that $c_{44}$ is replaced by 
its $z=0$ value Eq.\ (\ref{c44b}), i.e. the pinning and Bragg glass 
lengths are reduced.
The main result is then that the fluctuations in $u_T(r)$ behave with 
an effective $c_{44}$ which is large when $q<1/\lambda_c^R$
(Eq.\ (\ref{c44a})), i.e. for domain sizes $R_{BG}>\lambda_c^R$, while 
for $z=0$ $c_{44}$ is reduced (Eq.\ (\ref{c44b})). In the anharmonic 
region below decoupling (see Fig. 1) where $R_{BG}<\lambda_c^R$ the 
full form of Eq.\ (\ref{fluc}) is required to interpolate between 
these limits; this form avoids the ill-defined harmonic expansion in 
this regime.

Consider next the Josephson plasma frequency, given by 
$\omega_{pl}^2=(c^2/\epsilon_0\lambda_c^2)\langle \cos \tilde{b}_n(r)\rangle$
where $\epsilon_0$ is the dielectric constant
\cite{Matsuda,Koshelev}. The average in $\langle \cos \tilde{b}_n(r)\rangle$
 is on both thermal fluctuations and 
disorder and can yield significant information on the phase diagram. As 
shown by Koshelev \cite{Koshelev} the local $\langle \cos 
\tilde{b}_n(r)\rangle$ is finite even at high temperatures, e.g. above 
the decoupling transition. A high temperature 
expansion yields \cite{Koshelev} 
$\langle \cos \tilde{b}_n(r)\rangle = (E_J/2T)\int d^2r\exp [-A(r)]$
where $A(r)=\sum_{{\bf q},k}(1-\cos {\bf q}\cdot {\bf 
r})\langle |\tilde{b}^{\alpha}({\bf q},k)|^2\rangle$. The solution with 
disorder \cite{Horovitz} yields (up to a $\ln B$ dependence)
$A(r)=B(T+T_0)q_u^2r^2/(2B_0T_0)$ for  
$r<1/q_u$ where $q_u=2\ln ^{1/2}(a/d) /\lambda_{ab}$ while 
$A(r)\sim \ln q_u r$ or $\sim r$ for larger $r$. The $r$ integration is 
dominated by the short $r$ correlation which yields
\begin{equation}
\langle \cos \tilde{b}_n(r)\rangle \approx\frac{\pi E_J\lambda_{ab}^2}
{2\ln (a/d)}\cdot \frac{B_0T_0}{BT(T+T_0)} \label{cos}
\end{equation}
A $1/BT$ dependence 
has been obtained by Koshelev \cite{Koshelev} with a weakly 
temperature dependent prefactor for an XY model, i.e. infinite 
$\lambda_{ab}$. Data on BSCCO \cite{Matsuda} has shown that
$\langle \cos \tilde{b}_n(r)\rangle \sim B^{-0.8}T^{-1}$ in reasonable 
agreement with the $1/BT$ form. The present result shows that in fact the 
$1/BT$ form is valid in the disorder dominated regime, i.e $T<T_0$, 
though in general the fluctuation term yields 
$\omega_{pl}^2 \sim 1/[BT(T+T_0)]$.

Using the RSB solution, it can be shown that the Josephson glass 
parameter contributes 
a negative term to $\langle \cos \tilde{b}_n(r)\rangle$ so that 
$\omega_{pl}$ is reduced, while the Josephson coupling contributes a 
positive $\approx z/z_{bare}$ term which vanishes at 
decoupling. These are mean field results to which 
fluctuation terms, as Eq.\ (\ref{cos}), should be added. 
The recent data on BSCCO \cite{Shibauchi,Matsuda2} is consistent with
these results, i.e. a drop at the second peak transition followed by a
field dependent fluctuation term at higher fields.

In conclusion, it is shown that a decoupling transition leads to an 
apparent reduction in $c_{44}$ within an anharmonic region where the 
harmonic expansion fails. The proper 
interpolation across the anharmonic region is achieved by Eq.\ 
(\ref{fluc}). The reduction in $c_{44}$, the resulting reduction in 
domain sizes and the enhanced $j_c$ account for the hallmark feature 
of the second peak transition. Furthermore, $B_0$ being weakly $T$ 
dependent and decreasing with disorder \cite{Khaykovich1}, as well as the
Josephson plasma resonance data \cite{Shibauchi,Matsuda2}, lend
substantial support for the identification of the second peak transition
as a disorder induced decoupling.

\vspace{10mm}
{\bf Acknowledgments}: 
I thank  P. Le Doussal, T. Giamarchi, T. Natterman, G. Blatter and V. B.
Geshkenbein for most valuable discussions.
This research was supported by THE ISRAEL SCIENCE FOUNDATION founded 
by the Israel Academy of Sciences and Humanities.


\begin{references}
\bibitem{Blatter} For a review see G. Blatter {\em et al.}, Rev. Mod. Phys.
{\bf 66}, 1125 (1995).
\bibitem{LO} A. I. Larkin and  Y. N. Ovchinikov, J. Low Temp. Phys. 
{\bf 34}, 409 (1979).
\bibitem{Giamarchi} T. Giamarchi and  P. Le Doussal,
 Phys. Rev. B {\bf 52}, 1242 (1995).
\bibitem{Natterman} T. Natterman, Phys. Rev. Lett. {\bf 64}, 2454 
(1990); J. Kierfeld, T. Nattermann and  T. Hwa, 
Phys. Rev. B {\bf 55}, 626 (1997) .
\bibitem{Kes} For a review see P. H. Kes, J. Phys. I France {\bf 6}, 
2327 (1996).
\bibitem{Khaykovich1} B. Khaykovich et al., Phys. Rev. B, {\bf 56}, 
R517 (1997); Czech. J. Phys. {\bf 46-S6}, 3218 (1996).
\bibitem{Khaykovich2} B. Khaykovich et al., Phys. Rev. Lett. {\bf 76}, 
2555 (1996).
\bibitem{Deligiannis} K. Deligiannis et al.,  Phys. Rev. Lett. 
{\bf 79}, 2121 (1997).
\bibitem{Higgins} M. J. Higgins and S. Bhattacharya, Physica C {\bf 232}, 
232 (1996).
\bibitem{Marley} A. Marley et al.,  Phys. Rev. Lett. {\bf 74}, 
3029 (1995).
\bibitem{Bruynseraede} Y. Bruynseraede et al., Phys. Scr. {\bf T42}, 
37 (1992).
\bibitem{Otterlo} A. van Otterlo, R. T. Scalettar and G. T. Zimanyi,
cond-mat/9803021. 
\bibitem{Forgan}  E. M. Forgan et al., Czech. J. Phys. {\bf 46}-suppl. 
S3, 1571 (1996); S. L. Lloyd {\em et al.} (to be published).
\bibitem{Fuchs1} D. T. Fuchs et al., Nature {\bf 391}, 373 (1998).
\bibitem{Fuchs2} D. T. Fuchs et al., Phys. Rev. Lett. {\bf 80}, 4971 
(1998).
\bibitem{Horovitz} B. Horovitz and T. R. Goldin, Phys. Rev. Lett. {\bf 
80}, 1734 (1998).
\bibitem{Matsuda} Y. Matsuda et al., Phys. Rev. Lett. {\bf 78}, 1972 
(1997).
\bibitem{Koshelev} A. E. Koshelev, Phys. Rev. Lett. {\bf 77}, 3901 
(1996).
\bibitem{Shibauchi} T. Shibauchi et al., Phys. Rev. Lett. {\bf 83}, 
1010 (1999). 
\bibitem{Matsuda2} Y. Matsuda, private communication.
\bibitem{Dodgson} M. J. W. Dodgson, V. B. Geshkenbein and G. Blatter,
cond-mat/9902244.
\bibitem{Frey} E. Frey, D. R. Nelson and D. S. Fisher, Phys. Rev. B{\bf
49}, 9723 (1994).
\bibitem{Marchetti} M. C. Marchetti and L. Radzihovsky, Phys. Rev. B {\bf
59}, 12001 (1999).
\bibitem{Glazman} L. I. Glazman and  A. E. Koshelev, Physica 
C {\bf 173}, 180 (1991)
\bibitem{Daemen} L. L. Daemen, L. N. Bulaevskii, M. P. Maley and 
 J. Y. Coulter,  Phys. Rev. Lett. {\bf 70}, 1167 (1993).
\bibitem{Sudbo} A. Sudb{\o} and E. H. Brandt, Phys. Rev. Lett. {\bf 66}, 
1781 (1996).
\bibitem{GK} L. I. Glazman and A. E. Koshelev, Phys.Rev. B {\bf 
43}, 2835 (1991).
\bibitem{Goldin} T. R. Goldin and B. Horovitz, Phys. Rev. B {\bf 58}, 
9524 (1998).
\bibitem{Yeshurun} Y. Yeshurun, N. Bontemps, L. Burlachkov and A. 
Kapitulnik, Phys. Rev. B {\bf 49}, 1548 (1994).
\bibitem{Horovitz2} B. Horovitz, Phys. Rev. B {\bf 47}, 5947 (1993).

\end{references}
\end{document}